\documentclass{ws-procs9x6}
\newcommand{\be}{\begin{equation}}
\newcommand{\ee}{\end{equation}}
\newcommand{\<}{\langle}
\renewcommand{\>}{\rangle}
\begin{document}

\title{Composite bosons and quasiparticles  in a number conserving approach }

\author{ F.Palumbo  }

\address{INFN-Laboratori Nazionali di Frascati,\\
Frascati, P.O. Box 13, 00044, Italy\\
E-mail: Fabrizio.Palumbo@lnf.infn.it}

\begin{abstract}

I recently proposed a method of  bosonization 
valid for systems of an even number of fermions whose partition function is dominated at low energy by 
bosonic composites.  This method respects all symmetries, in particular fermion number conservation.
I extend it to treat odd systems  and excitations involving unpaired fermions.

\end{abstract}

\keywords{ Composite bosons and quasiparticles in superconducting systems}

\bodymatter

\section{Introduction}\label{aba:sec1}

There are many ways to introduce composite bosons as elementaty degrees of freedom in fermion systems, but
most of them violate fermion number conservation, which is an important symmetry in atomic
nuclei and other finite systems~\cite{Duke}. I recently developed a method to bosonize even systems which respects all
symmetries, including fermion number conservation~\cite{Palu}. This method was applied to relativistic
field theories at zero~\cite{Cara} and nonvanishing~\cite{Palu1} fermion number.
I present here an application of this technique to nonrelativistic systems. In Section 2 I make an outline of the formalism,
giving in Section 3 its physical interpretation. In Section 4 I report the effective action of composite bosons and
finally in Section 5 the effective action of composite bosons plus unpaired fermions. 

\section{Outline of the approach}

Consider the partition function of a fermion system 
\be
Z = \mbox{Tr}^F \left\{ \exp \left[ - { 1 \over T} ( \hat{H}_F - \mu_F \, \hat{ n}_F) \right] \right\}
\ee
where $\mbox{Tr}^F$ is the trace over the Fock space,
$T$  the temperature,  $\mu_F$ the chemical potential and $\hat{ n}_F $ the fermion number operator.
 A sector of $n_F$ fermions can be selected by the constraint
\be
T \,{\partial \over \partial \mu_F} \ln Z = n_F.             \label{fermionnumber}
\ee
 A functional form of $Z$ can be found by performing the trace over coherent states
\begin{equation}
|c \rangle = \exp \left( \sum_m   \hat{c}_m^{\dagger}\,  c_m\right) |0 \rangle \,,
\end{equation}
where $m$ are the fermion quantum numbers, $ \hat{c}_m^{\dagger} $ their canonical creation operators and
 $c_m$ Grassmann
 variables. Coherent states satisfy the basic or defining equations
\begin{equation}
\hat{c}_m | c \rangle = c_m |c \rangle
\end{equation}
where $\hat{c}_m  $ are canonical destruction operators. In terms of these states I can write the identity in the 
fermion Fock space in the form
\begin{equation}
\mathcal{I} = \int  \prod_m  dc^*_m dc_m  \, \langle c| c \rangle^{-1} | c \rangle \langle c| \,.
\end{equation}
Using the above resolution of the identity
the partition function can be written
\be
Z = \mbox{Tr}^F \left\{ \mathcal{I} \exp \left[ - \tau (\hat{ H}_F - \mu_F \, \hat{ n}_F) \right] \right\}^{L_0}
\ee
where I introduced  Euclidean time spacing $\tau$ and number of temporal sites $L_0$ by setting
\be
T = { 1 \over \tau L_0}\,.
\ee
The trace can be evaluated exactly at the relevant order in $\tau$ with the result
\begin{equation}
Z = \int [dc^* dc]  \exp \left[- S_F( c^*, c) \right]\,.
\end{equation}
 $S_F$ is the fermion action in Euclidean time 
\be
S_F=\tau \sum_t \left\{ 
 c_t ^* \nabla_t c_{t-1}  + H_F (c_t^*,c_{t-1}) 
 - \mu_F  \,c_t^* c_{t-1} \right\} \,, \label{actelem}
\ee
with the discrete time derivative 
\be
\nabla_t \, f = { 1 \over \tau} \left( f_{t+1} - f_t \right)  \,.
\ee
In a superconducting system one has to perform a  Bogoliubov transformation which introduces  quasiparticles 
creation-annihilation operators
\begin{eqnarray}
\hat{\alpha}& =& R ^{{1\over 2}}
 ( \hat{c} - \mathcal{B} \, \hat{c}^{\dagger})
\nonumber\\
\hat{\alpha}^{\dagger} & =& (c^{\dagger} - \hat{c} \, \mathcal{B}^{\dagger} )
R ^{{1\over 2}}\,, 
\end{eqnarray}
where
\be
R = \left( 1 + \mathcal{B}  \, \mathcal{B}^{\dagger} \right)^{-1} \,.\label{R}
 \ee 
They satisfy canonical commutation relations for an arbitrary matrix $\mathcal{B} $. 
To write down the Bogoliubov transformation of coherent states I introduce some definitions. First I 
define the creation operator of a bosonic composite with quantum number $K$
 \be
{\hat B}^{\dagger}_K
= { 1\over 2\sqrt{\Omega}} \sum_{m_1,m_2} {\hat c}^{\dagger}_{m_1}\left( B^{\dagger}_K\right)_{m_1,m_2}
{\hat  c}^{\dagger}_{m_2}\,.
\ee
$B_K$ is the structure function of the composite and  $\Omega$  its index of nilpotency
(assumed to be independent of $K$), defined as the largest integer such that
\be
\left( \hat{B}_K \right)^{\Omega} \neq 0. 
\ee
Next I expand the matrix $\mathcal{B} $ on the  matrices $B_K$
\be
\mathcal{B} = { 1\over \sqrt{\Omega}}\sum_K b_K B_K^{\dagger} \equiv  
{ 1\over \sqrt{\Omega}} \,b \cdot B^{\dagger}\,.
\ee
Lastly I define new coherent states
\be
| \alpha, b\> =\exp\left( - \sum_m \alpha_m \,\hat{\alpha}_m^{\dagger}\right) |b\> 
\equiv \exp( - \alpha \cdot \hat{\alpha}^{\dagger}) |b\>
\ee
where
\be 
 |b\> = \exp ( b \cdot \hat{B}^{\dagger}) |0\>
 \ee
 is the quasiparticle vacuum.The 
Bogoliubov transformation changes the coherent states $|c\>$ according to
\be
\mathcal{ S}| c\> =  \<  b |  b\>^{-{1\over 2}}  | c, b\> 
\ee
because \cite{Palu1}
\be
\< c,b | c, b\>  = \exp(c^* c) \< b |  b\> \,.
\ee
I can now write a new expression of the identity in the fermion Fock space in terms of the
transformed coherent states
\be
\mathcal{I}( b^*, b) = \int [dc^* dc] 
 \< c, b | c, b\>^{-1}  | c, b\> \<c, b|
\ee
and use it to get a new expression of the partition function
\be
Z = \mbox{Tr}^F \left\{ \mathcal{I} (b^*, b) 
\exp \left[ - \tau (\hat{ H}_F - \mu_F \, \hat{ n}_F) \right] \right\}^{L_0}\,.
\ee
Evaluation of the trace yields an action of the system strictly equivalent to the original one, but whose
individual terms do not respect several symmetries, in particular fermion number conservation. 

There are many cases in which  all symmetries can be enforced 
 term by term  proceeding in a slightly different way. I introduce at each Euclidean time an independent 
 Bogoliubov transformation $\mathcal{S}_t$. I do this by letting the expansion coefficients $b_K$
 to depend on time, $b_{K} \rightarrow b_{K,t}$ while keeping the basis matrices $B_K$ fixed.
Since nothing depends on the $\mathcal{B}$ and therefore on the expansion coefficients $b_{K,t}$ 
I can integrate over the latter ones in the partition function with an
arbitrary measure $d\mu(b^*,b)$
\be
Z = \int   d\mu(b^*,b) \, 
\mbox{Tr}^F \left\{ \prod_t \mathcal{I} (b^*, b_t) 
\exp \left[ - \tau (\hat{ H}_F - \mu_F \, \hat{ n}_F) \right] \right\}\,. \label{Z1}
\ee
Requiring the  variables $b^*,b$ to transform in the proper way all symmetries are restored.

Performing the trace, which can be done exactly~\cite{Palu1} at the relevant order in $\tau$, I get a new
functional form of the transfer matrix
\be
Z = \int   d\mu(b^*,b) \, 
  \prod_{m,t} \left [d \alpha_{m,t}^* d \alpha_{m,t} \right]   
\exp \left[ - S(b^*, b, \alpha^*,\alpha, B^{\dagger},B) \right]
\,. \label{Z2}
\ee

\section{Physical interpretation}

The last expression of the partition function contains, in addition to the Grassmann variables $\alpha^*, \alpha$,
the bosonic variables $b^*,b$. Is it possible to associate  these variables to physical bosons?

To answer this question let us analyze the nature of the states $| \alpha, b\> $. They are constructed in terms  of quasiparticles 
 and composite boson creation  operators
$\hat{\alpha}^{\dagger},{\hat B}_K^{\dagger}$.
I call these states  coherent  because they are obtained by a unitary transformation from truly coherent
states of fermions and they share with coherent states of elementary bosons
the property of a fixed phase relation among  components with different number of composites. 
But the basic property of coherent states cannot be fulfilled
\begin{equation}
\hat{B}_K | b \rangle \neq \, b_K | b\rangle\  \,.
\end{equation}
This is a consequence of the composites commutation relations, which are not canonical
\begin{equation}
\left[ \hat{B}_J ,   \hat{B}^{\dagger}_K  \right] = { 1 \over  2 \, \Omega}
\,  \mbox{Tr} \, (B_J  B^{\dagger}_K) 
- { 1 \over \Omega}\, {\hat c}^{\dagger}B^{\dagger}_K B_J \,{\hat  c }\,. \label{comm}
\end{equation}
In  states with a number of composites  $ n << \Omega$, the above equations can   approximate the canonical ones
by an appropriate normalization of the structure functions provided they are sufficiently smooth.
Indeed in such a case the last  term is of order $n / \Omega $. But in  states with  $n  \sim \Omega$ this
is impossible even with an absolute freedom about the form of the structure functions (which are instead determined by
the dynamics). The best we can do~\cite{Palu} is to satisfy them for states with $n+k$ composites, for fixed $n\sim \Omega$ and
 $|k| << \Omega$. 
 The above considerations can be made more precise, also taking quasiparticles into account, by showing
 that under the above conditions and for a number of quasiparticles much smaller than $\Omega$
 \begin{eqnarray}
&& \< (\hat{\alpha}_{m_1})^{r_1}... (\hat{\alpha}_{m_n})^{r_n}
 (\hat{b}_{K_1})^{r_{K_1}}...(\hat{b}_{K_t})^{r_{K_t}}| \int \prod_{K,t}
 \left[{db_{K,t}^* db_{K,t} \over 2 \pi i} \right]  \mathcal{I}(b^*,b)|
 \nonumber\\
 && 
  (\hat{\alpha}_{m_1}^{\dagger})^{s_1}... (\hat{\alpha}_{m_n}^{\dagger})^{s_n} 
 ( \hat{b}_{K_1})^{s_{K_1}}...   ( \hat{b}_{K_t})^{s_{K_t}}\>  \sim
 \delta_{r_1,s_1}...\delta_{r_t,s_t} \,.
 \end{eqnarray}
 {\it Only for such states can the holomorphic variables $b_K^*, b_K$  be interpreted as the fields of
  bosons with quantum numbers "K" in the presence of fermions with quantum numbers "m" associated 
  to the Grassmann variables $\alpha_m$}.
  Restricting myself to such states  I can assume the measure
  \be
 d\mu(b^*,b)=\prod_{K,t}\left[{db_{K,t}^* db_{K,t} \over 2 \pi i} \right]  \,.
  \ee
   The property
\be
\hat{\alpha} |b\>=0
\ee
ensures that there is no double counting: There are no quasiparticles in the bosonic composites.

What to do with states which do not satisfy the above requirements? We should integrate them out (which would
require a different choice of the measure) leaving only states 
which have a physical interpretation. But in some cases we can reasonably assume that we can ignore them,
in the spirit of a variational calculation.

\section{The effective action of composite bosons}

In a system of fermions whose low energy excitations are dominated by fermion composites  I can restrict the trace to 
these composites neglecting the quasiparticles. The restricted partition function can be written
 \be
Z_C = \mbox{Tr}^F \left\{ {\mathcal P}_C \exp \left[ - { 1 \over T} (\hat {H}_{F}- \mu_{F}  
\, \hat{ n}_{F})  \right] \right\}
\ee
where ${\mathcal P}_C$ is a projection operator in the subspace of the composites for which I assume the 
approximate form
\be 
{\mathcal P}_C = \int {db^* \,d b \over 2 \pi i  } \, 
\< b| b\>^{-1} |b\>\<b|   \,. \label{Poperator}
\ee
 To proceed further I write the fermion Hamiltonian in the form
\be
H_{F} =  \mu_{F}  
\, \hat{ n}_{F} + {\hat  c}^{\dagger} h_0 \, {\hat c} -
\sum_K  g_K  \, { 1\over 2} \, {\hat c}^{\dagger} F_K^{\dagger}{\hat c}^{\dagger} \, 
{ 1\over 2}\, {\hat c} \, F_K \,{\hat  c}. \label{Hami}
\ee
 The one-body term includes the single-particle energy 
with matrix $e$, the fermion chemical potential $\mu_{F}$ and any single-particle interaction with external fields
included in the matrix ${\cal M}$  
\be
h_0= e - \mu_{F}+ {\cal M} .
\ee
The matrices $F_K $ are the form factors of the potential, normalized according to
\be
 \mbox{tr} ( F_{K_1}^{\dagger} F_{K_2}) = 2 \, \Omega \, \delta_{K_1 K_2}.  \label{potnormal}
\ee
Now the trace in the partition function can be evaluated ~\cite{Palu}  yielding its functional form 
\be
Z_C=\int \left[ { d b^* d b \over 2 \pi i} \right]\exp \left( - S_C(b^*, b, B^{\dagger} , B) \right)
\ee
in which
 \begin{eqnarray}
& & S_C( b^*, b, B^{\dagger}, B )=\tau \sum_t { 1 \over 2} \mbox{tr} \left\{ { 1\over  \tau}   \ln \left[
1\!\!1 + 
 \tau \, R" \, {\mathcal B}^{\dagger} \nabla_t {\mathcal B} \right] \right.
\nonumber\\
& & \left.
+   2  \, R"   \,{\mathcal B}^{\dagger}   h \, {\mathcal B}    - 
 \sum_K g_K  \left[ (R" -1) \,  F^{\dagger}_K F_K  \right. \right.
\nonumber\\
& & \left. \left.
 +  (R" \, {\mathcal B}^{\dagger}  F_K^{\dagger}) \,
{ 1 \over 2} \mbox{tr}(R" \, F_K {\mathcal B}) 
-   R" \, {\mathcal B}^{\dagger} F_K^{\dagger} R" \, F_K {\mathcal B}  \right] 
 \right\} \,.
\label{bosaction}  
\end{eqnarray}
In this expression "$\mbox{tr}$" is the trace over fermion quantum numbers,
\be
h = h_0 -  \sum_K g_K   F_K^{\dagger} F_K  
\ee
and the variables $b^*,b$ must  be understood at times $t,t-1$ respectively, unless otherwise specified. So, for instance
\be
R" \rightarrow  \left( 1 + {\mathcal B}_t^{\dagger} {\mathcal B}_{t-1} \right)^{-1}  \,.
\ee
The latter function should be distinguished from the function R introduced in Eq.(\ref{R})
\be
R =  \left( 1 + {\mathcal B} \, {\mathcal B}^{\dagger} \right)^{-1}  
\rightarrow  \left( 1 + {\mathcal B}_{t-1} {\mathcal B}_t^{\dagger} \right)^{-1}  \equiv  R_{t,t-1}\,.
\ee
The  derivation of the above functional form of the partition function  is based only on 
the physical assumption of boson dominance and the 
approximation adopted 
for ${\mathcal P}_C $. {\it  The action $S_C$ is a functional of the structure matrices $B_K$ 
which can be determined by a variational calculation}.

In many-body physics it is often preferred the Hamiltonian formalism. The Hamiltonian of the effective bosons, $H_B$,
cannot be read directly from the effective action, because $S_C  $ does not have the form
of an action of elementary bosons. Indeed it contains anomalous time derivative terms, anomalous couplings of the
chemical potential and nonpolynomial interactions, which are all features of compositeness. Therefore it has
been necessary to devise an appropriate procedure to derive $H_B$, which is  given\cite{Palu} in terms of 
{\it canonical} boson operators
 ${\hat b}^{\dagger}, {\hat b}$, 
(not to be confused with the composite operators  ${\hat B}^{\dagger}, {\hat B}$), so that
\be
Z_C = \mbox{Tr}^B \left[ -{ 1 \over T} ({\hat H}_B - \mu_B \, \hat{ n}_B) \right].
\ee
${Tr}^B $ is the trace on the boson Fock space, $\mu_B$ is the boson chemical potential and $\hat{ n}_B $ the boson number operator.

${\hat H}_{B}$ has a closed form but, for a practical use, it is necessary to perform an expansion 
in  inverse powers of the index of nilpotency $\Omega$.

\section{The effective action of composite bosons plus quasiparticles}

Finally I consider the case in which one needs to retain, in addition to composite bosons, also
 quasiparticles. To find the total effective action I first observe that the new coherent states can be rewritten
\be
 | \alpha, b\> =  \exp \left[ -{1\over 2} \alpha \,(R^*)^{{ 1\over2}} 
\mathcal{B}^{\dagger} \, R^{{ 1\over2}} \alpha \right]
|\alpha \,, b \>\>
\ee
where I introduced the definition
\be
|\alpha \,, b \>\> =
 \exp \left(  \hat{c}^{\dagger}  \, R^{-{1 \over 2}}\alpha \right)|b\> \,.
\ee
Notice that in the latter states there appear the original fermion operators, not the  quasiparticle ones.

According to Eqs. (\ref{Z1},\ref{Z2}) the action is given in terms of the ratios
\be
\< \alpha_1 b_1 | \alpha_2 b_2 \> ^{-1} \< \alpha_1 b_1 | \exp(-\tau \hat{H})| \alpha_2 b_2 \> \,. \label{ratio}
\ee 
The matrix element of any operator $ \mathcal{O}$ factorizes according to
\begin{eqnarray}
 \< \alpha_1, b_1 | \mathcal{O}| \alpha_2, b_2\> &=&
 \exp \left( - { 1\over 2} \alpha_1^* R_{11}^{{1\over 2}} \mathcal{B}_1
 (\mbox{R}^T_{11})^{{1\over 2}} \alpha_1^* 
  -   { 1\over 2} \alpha_2  (R_{22}^*)^{{1\over 2}} \mathcal{B}_2^{\dagger} \,
 R_{22}^{{1\over 2}} \alpha_2   \right)
 \nonumber\\
 & & \times \<\< \alpha_1, b_1 |\mathcal{O}| \alpha_2, b_2\> \> \,.
 \end{eqnarray}
 The first factor cancels out in the ratios (\ref{ratio}) and the second factor is given by
 \be
\<\< \alpha_1, b_1 | \exp (- \tau \hat{H}_F ) | \alpha_2, b_2\>\> =
\<  b_1 | \exp (- \tau \hat{H}_F ) | b_2\>
\exp ( - \mathcal{s} -  \tau H_{qp}) \,.
\ee
The first matrix element yields the composite boson term $S_C$ of the action reported in the previous Section,
while the second one is the quasiparticle contribution. To shorten its expression I define the Grassmann variables
\begin{eqnarray}
w^* &=& R^T \left[ \left(R^T\right)^{-{ 1\over 2}}\alpha^*
-\mathcal{B}^{\dagger} \, R^{-{ 1\over2}} \,\alpha \right]
\nonumber\\
w &= &R \left[  R^{-{ 1\over2}}\alpha + 
\mathcal{B} \left(R^T\right)^{-{ 1\over 2}} \,\alpha^*\right] \,.
\end{eqnarray}
In terms of these variables
\begin{eqnarray}
\mathcal{s} &= & -w^*w - { 1\over 2} w^* {\mathcal B} \,  w^*
- { 1\over 2} w \,  {\mathcal B}^{\dagger} w
\nonumber\\
H_{qp} &=&   w^* \left( h + \sum_K g_K  F_K^{\dagger} \, R^T  F_K \right)  \, w 
 - { 1\over 4} \sum_K g_K  \,   \, \mbox{tr }\left[ \left({\mathcal B}^{\dagger} \,
 R \, F_K^{\dagger}\right) \right.
  \nonumber\\
&  \times &  \left. 
 w \, F_K  \, w
 + \left( F_K   \, R \, {\mathcal B} \right)     w^* F_K w^*   \right]
  - \sum_K  g_K \,
{ 1\over 2} w^* F_K^{\dagger} w^* \,{ 1\over 2} w F_K w \,.
 \end{eqnarray}
Now I restrict myself to the case of a number of quasiparticles much smaller than $\Omega$.  Then assuming
$ \nabla_t b_K\sim \nabla_t b_K^*\sim 1$ the noncanonical temporal terms in the action are
of order $\Omega^{-{1\over 2}}$
\be
 { 1\over 2} w_t^*( {\mathcal B}_{t-1}- 
{\mathcal B}_t) w_t^*
+ { 1\over 2} w_{t-1} ( {\mathcal B}_t^{\dagger} - 
{\mathcal B}_{t-1}^{\dagger}) w_{t-1} = O(\Omega^{-{1\over 2}})
 \ee
 because ${\mathcal B} = \Omega^{-{ 1\over 2}} b \cdot B^{\dagger}$ and $b_2-b_1 = O(1)$.
To this approximation the temporal terms in the action become $ \alpha_1^*( \alpha_1- \alpha_2)$,
which are canonical,  and therefore the total Hamiltonian can be written in operator form
\be
\hat{H} = \hat{H}_B + \hat{H}_{qp}
\ee
where
\begin{eqnarray}
\hat{H}_{qp} &=& :\, \left\{-\hat{ w}^{\dagger} ( h + \sum_K g_K  F_K^{\dagger} R^T F_K)  \hat{w} 
 - { 1\over 4} \sum_K g_K    \, \mbox{tr } \left[ \left({\mathcal B}^{\dagger} \,
 R F_K^{\dagger}\right) \right. \right.
  \nonumber\\
&\times &  \left. \left.
 \hat{w}  F_K \hat{w} 
 + \left( F_K   \, R \, {\mathcal B}\right) \, \hat{ w}^{\dagger} F_K^{\dagger}\hat{ w}^{\dagger} \right]
 - \sum_K  g_K \,
{ 1\over 2} \hat{ w}^{\dagger} F_K^{\dagger}\hat{ w}^{\dagger} \,{ 1\over 2} \hat{w}  F_K \hat{w}  \right\}:
\end{eqnarray}
In this expression   the operators $\hat{w}^{\dagger},\hat{w} $ must be regarded as functions of  
the boson and quasiparticle operators 
$\hat{b}^{\dagger},\hat{b},\hat{\alpha}^{\dagger},\hat{\alpha}$ and the colons mean normal
order with respect to the latters.
  
  I do not have the space to discuss the properties of the Hamiltonian I have presented. I only make the following
  observations
  
  i) it respects term by term all fermion symmetries, in particular fermion number conservation, as it can be seen remembering
  that the bosonic operators $\hat{b}^{\dagger}$ have fermion number 2
  
  ii) it contains quadratic quasiparticles terms of the type $ \hat{\alpha}^{\dagger}\hat{\alpha}^{\dagger}$ and
  $ \hat{\alpha}\hat{\alpha}$, similar to the
  "dangerous" terms of Bogoliubov. These terms however do not break fermion number conservation because they are accompanied 
 by the operators $\hat{b}, \hat{b}^{\dagger}$
    respectively. Using  the variational equations satisfied by the structure functions of the composites, Eq.(81) of Ref. 2,
    one can see that the coefficients of these terms vanish and the term involving the operator   
    $  \hat{\alpha}^{\dagger}\hat{\alpha}$ takes the usual form.
    Therefore the difference with respect to the standard Bogoliubov transformation is that the quartic quasiparticle term
    involves interactions whith composite bosons.
    
\section{Acknowledgments }
This work was partly supported by PRIN 2006021029 "Complex problems in statistical mechanics and
field theory" of Ministero dell'Universita' e della Ricerca Scientifica.

\end{document}